\documentstyle[psfig]{aipproc}

\def\HI{H\,{\sc i}}
\def\HII{H\,{\sc ii}}
\def\fdg{\hbox{$.\!\!^\circ$}}
\def\la{\ifmmode\stackrel{<}{_{\sim}}\else$\stackrel{<}{_{\sim}}$\fi} 
\def\ga{\ifmmode\stackrel{>}{_{\sim}}\else$\stackrel{>}{_{\sim}}$\fi} 
\def\farcm{\hbox{$.\mkern-4mu^\prime$}}

\begin{document}

\title{High Resolution Polarimetry of the Inner Galaxy}

\author{Bryan M. Gaensler$^1$, J. M. Dickey$^2$,
N. M. McClure-Griffiths$^3$, \\ N. S. Bizunok$^4$, A. J. Green$^5$}
\address{$^1$Harvard-Smithsonian Center for Astrophysics,
  Cambridge MA 02138, USA\\
$^2$University of Minnesota, Minnesota MN 55455, USA\\
$^3$Australia Telescope National Facility, Epping NSW 1710, Australia \\
$^4$Boston University, Boston MA 02215, USA \\
$^5$University of Sydney, NSW 2006, Australia}

\maketitle

\begin{abstract}

We present our results from the Southern Galactic Plane Survey, an effort
to map the fourth quadrant of the Milky Way in linear polarization at
a frequency of 1.4 GHz and at a resolution of 1--2~arcmin. These data
are a powerful probe of both the turbulence and large-scale structure of
magneto-ionic gas, and have revealed a variety of new features in the
interstellar medium.

\end{abstract}

\date{\today}

\maketitle

\section*{Introduction}

The Milky Way was the first celestial radio source discovered, and was
subsequently one of the first sources to be detected in linear
polarization.  There are two sources of this polarization:  discrete
objects such as supernova remnants (SNRs), and a diffuse polarized
background produced by the relativistic component of the interstellar
medium (ISM). All of this emission undergoes Faraday rotation as it
propagates towards us, either in the source itself or in intervening
material.  With sufficiently high angular and frequency resolution, we
can use the properties of this polarized emission to map out the
distribution of ionized gas and magnetic fields in individual sources
and in the ambient ISM. Only recently have instruments and
techniques advanced to a point where such studies are feasible 
\cite{gld+99,hkd00,wdj+93}.

Motivated by the spectacular single-dish polarization surveys of Duncan
et al \cite{dhjs97,drrf99}, 
we have made polarimetric images of the entire fourth
quadrant of the Galaxy with the Australia Telescope Compact Array
(ATCA). These data have been taken as part of the Southern Galactic
Plane Survey (SGPS; \cite{mgd+01}).  While the primary focus of the SGPS is to
study the Galactic distribution of \HI, the ATCA simultaneously
receives full polarimetric continuum data, which have allowed us to map
out the distribution of linearly polarized emission in the survey
region.

While the full survey has now been completed, 
a detailed analysis has only been carried out
on a 28-deg$^2$ test region, 
covering the range
$325\fdg5 < l < 332\fdg5$, $-0\fdg5 < b < +3\fdg5$. We here summarize
the main results of this analysis; this work is described in more
detail by Gaensler et al \cite{gdm+00}.

\section*{Observations and Reduction}

The ATCA is a 6-element synthesis telescope,
located near Narrabri, NSW, Australia. 
Observations of the test region of the
SGPS were carried out in nine observing runs in 1997 and 1998,
and comprised 190 separate telescope pointings (see \cite{mgd+01} for details).
Data were recorded in nine spectral channels
spread across 96~MHz of bandwidth and centered at a frequency
of 1384~MHz. The two sources
MRC~B1438--481 and
MRC~B1613--586 were observed over a wide range in parallactic angle
in order to solve for the instrumental polarization characteristics of 
each antenna \cite{shb96}. For
each spectral channel, images of the field in Stokes $I$, $Q$, $U$ and
$V$ were deconvolved
jointly using the maximum entropy algorithm {\tt PMOSMEM}\ \cite{sbd99} and
then smoothed to a resolution of $\sim$1~arcmin. The
final sensitivity in each image is $\la0.5$~mJy~beam$^{-1}$.

Images of linearly polarized intensity, $L = (Q^2 +
U^2)^{1/2}$, linearly polarized position angle, $\Theta = \frac{1}{2}
\tan^{-1}(U/Q)$, and uncertainty in position angle, $\Delta\Theta =
\sigma_{Q, U}/2L$, were then formed from each pair of $Q$ and $U$
images. The nine $L$ maps (one per spectral channel) were then averaged
together to make a final image of $L$ for the entire test region, while
the nine $\Theta$ and $\Delta\Theta$ maps were used to derive an image
of the rotation measure (RM) over the field.

\section*{Results}

\begin{figure}
\centerline{\psfig{file=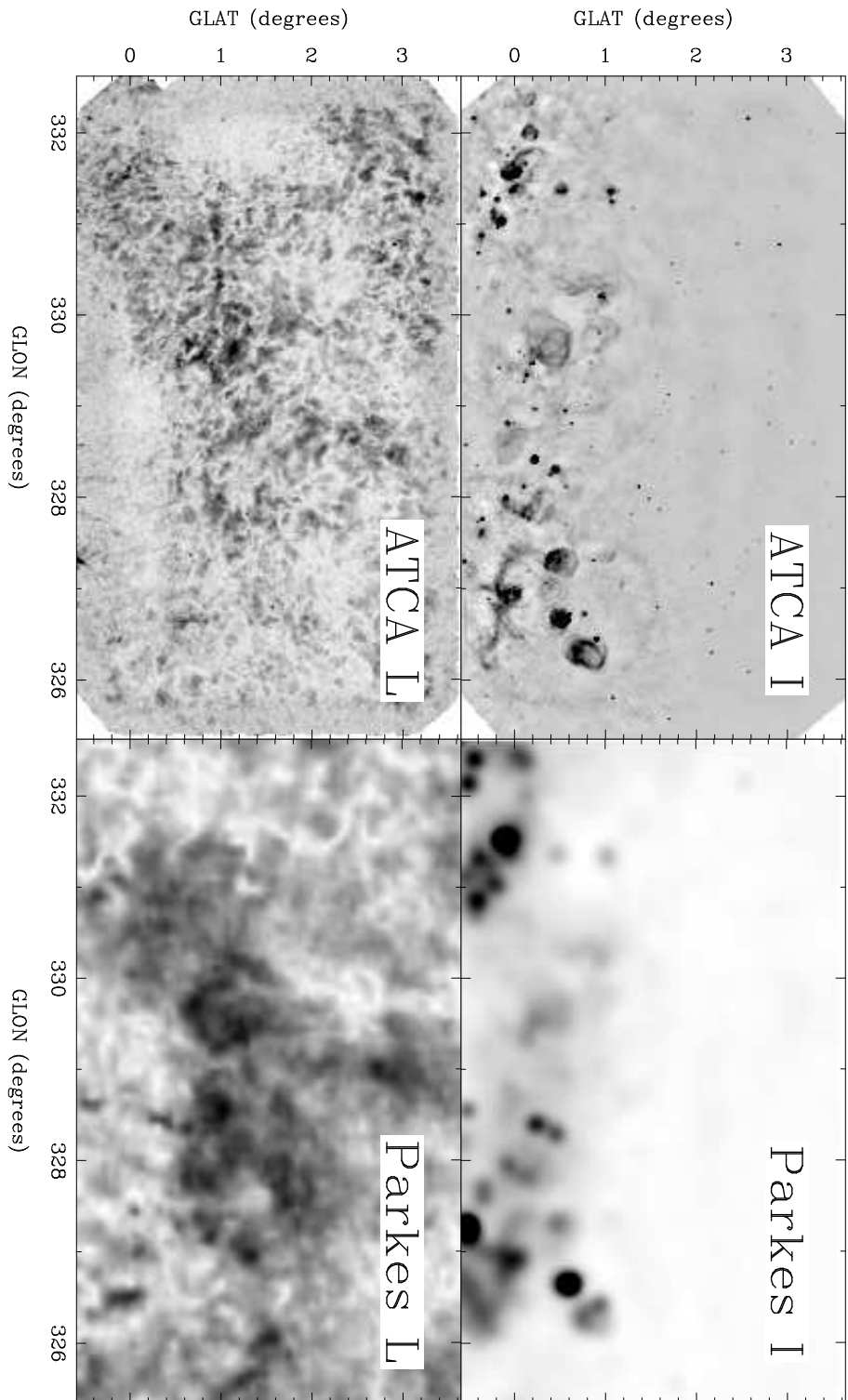,height=20cm,angle=180}}
\caption{Images of the SGPS test region with the ATCA
(1.4~GHz, 1~arcmin resolution) and Parkes (2.4~GHz, 10.4~arcmin),
in both total and linearly polarized intensity.}
\label{fig_field}
\end{figure}

In Figure~\ref{fig_field} we show images of $I$ and $L$ from both our
1.4-GHz ATCA observations and from the 2.4-GHz Parkes survey of Duncan
et al \cite{dhjs97} (resolution $10\farcm4$). The total intensity images show
the presence of SNRs and \HII\ regions (see \cite{mgd+01} for further
discussion).  Although the interferometric ATCA observations are not
sensitive to the diffuse emission seen by Parkes, it is clear that the
same features are present in both data-sets.

At first glance it seems that the $L$ images have very little in
common with the Stokes~$I$ emission.  In particular, the ATCA $L$
image is dominated by diffuse polarization spread all over the field
of view, composed of discrete patches separated by narrow ``canals''
of reduced polarization.  While none of this emission is correlated
with total intensity, there does seem to be a good match between the
brightest polarized regions of the ATCA and Parkes data, despite the
differing frequencies and resolutions of these data-sets.  Using the
images of $\Theta$ and $\Delta\Theta$, we can determine the variation of
polarization position angle with frequency wherever we detect polarized
emission. The resulting RMs are generally small and negative, with
a mean RM for the entire field of $-12.9\pm0.1$~rad~m$^{-2}$; 50\%
of the RMs have magnitudes smaller than $\pm$25~rad~m$^{-2}$ and 98\%
are smaller than $\pm$100~rad~m$^{-2}$.

The ATCA $L$ image reveals two
large voids of reduced polarization, each elliptical and several
degrees in extent. One void is centered on $(l, b) = (332\fdg4,
+1\fdg4)$ (``void 1'') and the other on $(328\fdg2,
-0\fdg5)$ (``void 2''); both voids are also seen
in the 2.4-GHz Parkes polarization map. The RMs around the edges
of these voids range up to $\pm400$~rad~m$^{-2}$, in distinction
to the low RMs seen over the rest of the field.

\begin{figure}
\centerline{\psfig{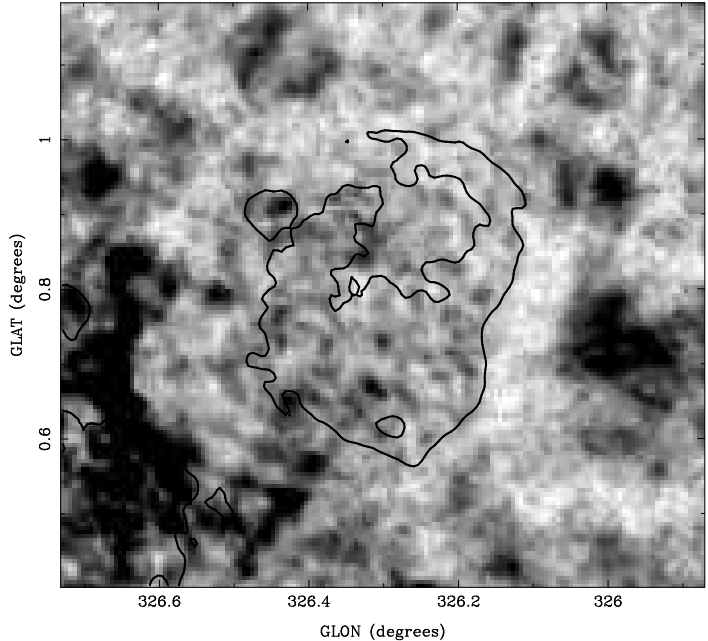}}
\caption{Polarized emission towards the \HII\ region RCW~94.  The
greyscale represents linearly polarized intensity, while the single
contour corresponds to total intensity emission from the same region at
the level of 45~mJy~beam$^{-1}$.}
\label{fig_rcw}
\end{figure}

A careful examination shows one marked correspondence between the ATCA
Stokes~$I$ and $L$ images:  at $(326.3, +0.8)$, the bright \HII\ region
RCW~94 shows reduced polarization towards its interior, and is further
surrounded by a halo in which no polarization at all is seen. This is
shown in more detail in Figure~\ref{fig_rcw}. 

Finally, of the numerous unresolved sources distributed across the field,
21 of these sources show detectable linear polarization. The RMs
for these sources fall in the range --1400~rad~m$^{-2}$
to +200~rad~m$^{-2}$. 

\section*{Discussion}

\subsection*{Diffuse Emission}

We first note that the incomplete $u-v$ coverage of an interferometer
affects images of polarization in complicated ways. While it is physically
required that $L \le I$, and we generally expect that structures seen
in $L$ might correspond to similar structures in $I$, neither situation
will be generally observed in interferometric data. This is because
an interferometer can not detect structures larger than a certain
size, corresponding to the closest spacings between its antenna
elements (in the case of the ATCA, this maximum scale
of $\sim35'$). A source larger than this maximum scale 
will not be seen in Stokes~I; if it is also a uniformly polarized source,
it will not be detected in polarization either. However,
magnetic field structure within the source, plus variations
in the Faraday rotation along different lines-of-sight,
can introduce power in Stokes $Q$ and $U$ on smaller scales,
to which the interferometer is sensitive. We thus
can often observe complicated structures in polarization
which have no counterpart in total intensity
\cite{gldt98,gld+99,hkd00,wdj+93}.

Clearly such an effect is occurring here, and is producing
virtually all the linear polarization seen in Figure~\ref{fig_field}. 
We can crudely divide up the diffuse polarization we see into two
components. 

The brightest polarization seen with the ATCA
matches well the bright polarized structures seen with Parkes. Since
the amount of Faraday-induced polarization is very strongly
dependent on both resolution and frequency, the fact that
two such disparate data-sets show similar structures implies
that these bright polarized structures are intrinsic
to the emitting regions. 
By comparing the RMs observed for this emission to those
observed for pulsars in this part of the sky, we can
conclude that the distance to this emission is in the range
1.3--4.5~kpc. The depolarizing effects of RCW~94
(discussed further below) imply that the polarized
emission is $>$3~kpc distant, while the lack of depolarization
against other \HII\ regions gives an upper limit of $6.5$~kpc.
Dickey \cite{dic97} has made \HI\ absorption measurements
against this emission to derive a lower limit on its distance
of 2~kpc. Taking into account all
these constraints, we argue that the mean distance
to the source of polarized emission is $3.5\pm1.0$~kpc,
corresponding to the Crux spiral arm of our Galaxy. 

The rest of the ATCA field is filled with fainter diffuse polarization,
which does not have any counterpart in the Parkes data. This
emission is best explained as being due to Faraday rotation in
foreground material. The RMs measured
for this emission imply that they are caused by foreground clouds of 
RM~$\sim 5$~rad~m$^{-2}$, consistent with the conclusions
made by Wieringa et al \cite{wdj+93}. 

\subsection*{Voids in Polarization}

To the best of our knowledge, voids in polarization such as those described
here have not been previously reported. There are two possible explanations
to account for these structures: either they represent regions where the
level of intrinsic polarization
is low, or they are the result of propagation through a foreground
object, whose properties have depolarized the emission at both 1.4 and
2.4~GHz. 

If the voids are intrinsic to the emitting regions, then the
distance of 3.5~kpc inferred above implies
that they are hundreds of parsecs across --- it is hard to see what
could produce such uniformly low polarized intensity across such large
regions. We thus think it unlikely that the voids are intrinsic
to the emitting regions.

We thus favor the possibility that the voids are caused by depolarizing
effects in foreground material. We have considered in detail
the various ways in which foreground Faraday rotation can 
produce the observed structure,
and can rule out bandwidth and gradient
depolarization as possible mechanisms (see \cite{gdm+00} for details).

The only remaining possibility is that depolarization in the voids is
due to beam depolarization, in which the RM varies randomly on small
scales.  We have developed a detailed model for ``void~1'' to confirm
this.  We consider void~1 to be a caused by a sphere of uniform
electron density $n_e$~cm$^{-3}$, centered on $(332\fdg5, +1\fdg2)$
with a radius of $1\fdg4$ and at a distance to us
of $d$~kpc. Within the sphere, we suppose that there are 
random and ordered components to the magnetic field, and
that these two components have identical amplitudes $B$~$\mu$G.  The ordered
component is uniformly oriented at an
angle $\theta$ to the line of sight.  We assume
that the random component is coherent
within individual cells of size $l$~pc, but that the orientation from
cell to cell is random.  Uniformly polarized rays which propagate
through a different series of cells will experience differing levels of
Faraday rotation, resulting in beam depolarization when averaged over
many different paths.

\begin{figure}
\centerline{\psfig{file=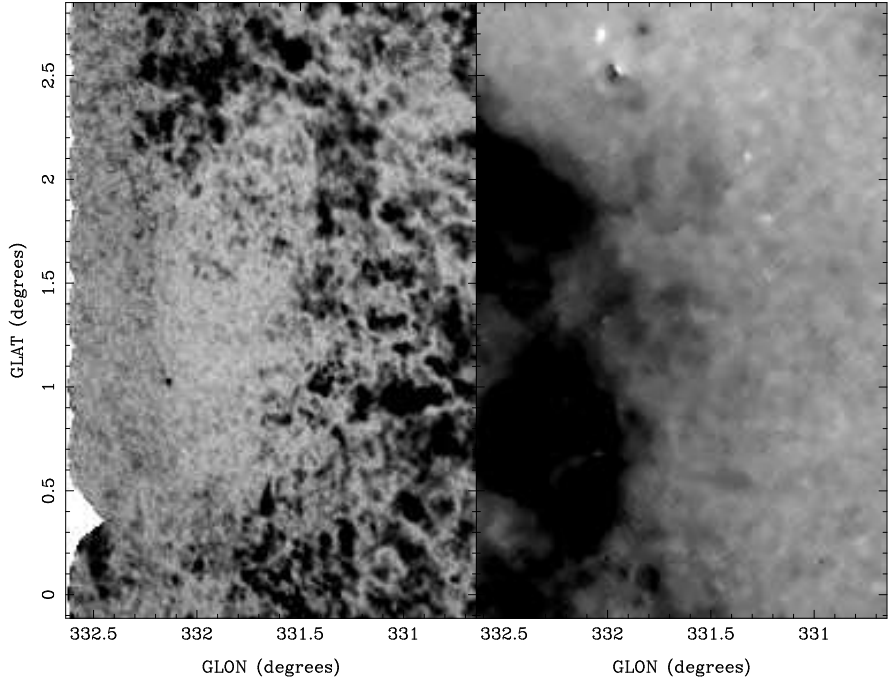,height=9cm,angle=270}}
\caption{Comparison of linear polarization (left, \protect\cite{gdm+00})
and H$\alpha$ emission (right, \protect\cite{gmrv01}) towards
void~1.}
\label{fig_halpha}
\end{figure}

By calculating the properties of the polarized signal which emerges
after propagating through this source, we find that we can account for
the observed properties of void~1 if $n_e \sim 20$~cm$^{-3}$, $B
\sim 5$~$\mu$G, $\theta \ga 80^\circ$, $d \sim 300$~pc and $l \sim
0.2$~pc (see \cite{gdm+00} for details). These properties
are consistent with those of an \HII\ region
of comparatively low emission measure. Indeed 
Figure~\ref{fig_halpha} demonstrates that
H$\alpha$ emission fills void~1,
its morphology and perimeter
matching exactly to that of the void.
It is interesting to note that the O9V star
HD~144695 is very close to the projected
center of void~1, and is at a distance of $300\pm160$~pc.
The radius of the Str\"{o}mgren sphere which this star
would produce is consistent with the extent of the void.
It is thus reasonable to propose that the star is embedded
in and powers the surrounding ionized bubble.

Two properties of the voids which our simple model cannot account for
are the requirement that the uniform component of the magnetic field be
largely oriented in the plane of the sky, but that we generally observe
coherent regions of large RM (of the order of a few hundred
rad~m$^{-2}$) around the edges of the voids. We suggest that both these
results can be explained by the field
geometry which arises during the expansion phase of an
\HII\ region as it interacts with surrounding material.
This produces a magnetic field perpendicular to the line
of sight over most of the void, but which is parallel to the line of
sight (and can thus potentially produce high RMs) around the perimeter.

\subsection*{Depolarization seen towards RCW~94}

The reduced polarization seen coincident with RCW~94 in
Figure~\ref{fig_rcw} presumably results
from beam depolarization, just as for the \HII\ region argued to
produce void~1. However, the effects of beam depolarization are
expected to be weakest around the edges of the source, and thus cannot
account for the halo of complete depolarization surrounding RCW~94.  We
rather account for this depolarization halo by requiring the electron
density to be approximately constant across RCW~94, but to fall off
rapidly beyond the boundaries of the source. This produces a sharp
gradient in RM around the edges of the source, resulting in complete
depolarization.

The presence of significant CO emission at the same position and
systemic velocity as for RCW~94 \cite{bact89} suggests that the
\HII\ region is interacting with a molecular cloud.  This possibility
is supported by \HI\ observations of the region, which show that RCW~94
is embedded in a shell of \HI\ emission, which is further surrounded by
a ring of decreased \HI\ emission \cite{mdg+00b}.  McClure-Griffiths
et al \cite{mdg+00b,mgd+01} argue that this structure in
\HI\ confirms that RCW~94 is embedded in a molecular cloud, the shell
of emission resulting from H$_2$ molecules dissociated by the
\HII\ region, and the surrounding region of reduced \HI\ corresponding
to regions of undisturbed molecular material.  Simulations of
\HII\ regions evolving within molecular clouds (\cite{rtf95} and
references therein) show that for certain forms of the density profile
within the parent cloud, the shock driven into the cloud by the
embedded expanding \HII\ region can produce a halo of partially ionized
material around the latter's perimeter, which would produce the
fall-off in $n_e$ required to produce the depolarization halo
observed. 

\subsection*{Point Sources}

With the exception of one source known to be a pulsar,
the polarized point sources in our field are presumably extragalactic, and
their RMs thus probe the entire line-of-sight through the
Galaxy. When combined with information from pulsar RMs,
we can use these data to constrain the geometry of
the overall Galactic magnetic field. So far
we have compared the RMs in our test region to
those expected for a bisymmetric spiral configuration,
and have found that pitch angles in the lower end
of the range allowed by pulsars ($p\sim-4.5^\circ$) are
favored \cite{dom+01}. We are in the process of carrying
out a more detailed study using the RMs of 163 background
sources from the entire SGPS, in which we are comparing
these measurements to the distributions expected for
a wider variety of geometries and model parameters
(Bizunok et al, in preparation).

\section*{Conclusions}

The ATCA's sensitivity, spatial resolution and spectral flexibility have
allowed us to study linear polarization and Faraday rotation from the
inner Galaxy in an unprecedented detail. Even though the test region
we have considered covers less than 7\% of the full survey, we have
been able to identify a variety of distinct polarimetric phenomena, and
have used these to map out both global and turbulent structures in the
magneto-ionized ISM. We anticipate that our analysis of the full SGPS
will result in a comprehensive study of magnetic fields and turbulence
in the inner Galaxy.

\begin{acknowledgments}

The Australia Telescope is funded by the Commonwealth of Australia for
operation as a National Facility managed by CSIRO.  H$\alpha$ data were
taken from the Southern H-Alpha Sky Survey Atlas (SHASSA), which is
supported by the National Science Foundation.  B.M.G. is supported by
a Clay Fellowship awarded by the Harvard-Smithsonian
Center for Astrophysics, while J.M.D.  acknowledges the support of NSF grant
AST-9732695 to the University of Minnesota.

\end{acknowledgments}

\clearpage

\bibliographystyle{aipproc}
\bibliography{journals,modrefs,psrrefs}

\end{document}